# All-Optical Excitable Spiking Laser Neuron in InP Generic Integration Technology

List of authors:

Lukas Puts[1,*]
Daan Lenstra[1]
Kevin A. Williams[1]
Weiming Yao[1]

List of Affiliations:

[1] Photonic Integration Group, Eindhoven Hendrik Casimir Institute (EHCI)
 Eindhoven University of Technology (TU/e), P.O. Box 513, 5600 MB, Eindhoven, the Netherlands

Contact information:

l.puts@tue.nl, ▬▬▬▬▬▬▬▬ (corresponding author)
d.lenstra@tue.nl
k.a.williams@tue.nl
w.yao@tue.nl

Keywords:

neuromorphic photonics, spiking neural network, excitability, photonic neuron, laser neuron, optical injection

**Abstract**

*Brain-inspired, neuromorphic devices implemented in integrated photonic hardware have attracted significant interest recently as part of efforts towards novel non-von Neumann computing paradigms that make use of the low loss, high-speed and parallel operations in optics. An all-optical spiking laser neuron fabricated on the indium-phosphide generic integration technology platform may be a practical alternative to other semi-integrated photonic and electronic-based spiking neuron implementations. Owing to the large number of predefined building blocks, a plethora of applications have benefitted already from the generic integration process. This technology platform has now been utilised for the first time to demonstrate an all-optical spiking laser neuron. This paper present and discusses the design and measurement of the ultra-fast and rich spiking dynamics in these devices. We show that under external pulse injection and operated slightly below the lasing threshold, the laser neuron exhibits an excitable mode, in addition to a self-spiking mode far above the threshold when no pulse is injected. In the excitable mode, the required injected pulse energy is much lower than that of the generated excited response, meeting an important requirement for neuron cascadability. In addition, we investigate excitability at different injection wavelengths below the lasing wavelength, as well as the ultra-fast temporal properties of the spiking response. All of the discussed characteristics point to the laser neuron being an important candidate for scaling up to future fully-connected, multi-wavelength all-optical photonic spiking neural networks in indium-phosphide generic integration technology.*

1. **Introduction**

Computing is indisputably a vital part of modern society. Soon after the invention of the integrated circuit in 1958, computers became ordinary. Already in 1945 – two decades before the invention of the integrated circuit – the Hungarian-American scientist John von Neumann led a group of scientist to propose a processing architecture for a digital computer. For decades, processors based on this architecture were very successful in providing computational power. In von Neumann's architecture, the processor fetches and operates instructions from memory on data sequentially, since the memory and data share the same physical bus. At high operating speed and data throughput, this becomes a processing bottleneck, which is known as the 'von Neumann bottleneck'.

The von Neumann bottleneck has detrimental effects on the efficiency of machine learning algorithms implemented on traditional processors, because machine learning models often rely on a fundamental operation in linear algebra: the vector-matrix multiplication. In the context of machine learning, this operation applies a weight matrix to an input vector. Inherently a parallel mathematical operation, this multiplication cannot be executed efficiently on a processor that works in a serial way. For more efficient machine learning implementations, a shift towards designing hardware that is processing data in a parallel way seems therefore imminent.

The deficiencies of traditional hardware has led to the interest in neuromorphic computation, the field of designing hardware architectures that are inspired by the human brain. At an estimated number of 100 billion neurons and 7000 synaptic connections per neuron[1], the human brain is a densely connected network that operates in a parallel way. Furthermore, with a power consumption of only 20 watts, the human brain is incredibly energy efficient.[2] One of the reasons for this is the sparsity and event-based signalling of this network; in a biological neuron, the action potential is generated only when a perturbation drives the neuron's membrane potential across a certain threshold, otherwise it remains in its quiescent state. This all-or-nothing non-linear response to a sufficiently strong input perturbation is referred to as 'excitability'.

Another important property of excitability in biological neurons is the refractory period. This refers to the recovery state right after the action potential generation, in which a second perturbation does not result in a response. Next to excitable neurons, a different class of neurons is capable of generating a sustained train of spikes, a phenomenon that is called 'tonic-spiking' or self-spiking, with frequencies ranging from 5 – 150 Hz.[3] Such dynamic effects significantly contribute to the large processing capacity and energy efficiency of spiking neurons.

Inspired by the brain's parallelism and efficiency, considerable efforts were made to realise hardware-based spiking neural networks (SNNs). Examples in electronics are IBM's TrueNorth, the University of Manchester's SpiNNaker, Intel's Loihi 1 and 2, and recently Innatera's T1.[4] Targeting different applications, from brain simulations using TrueNorth to low-power processing of sensor data using T1, these are all based on electronic SNNs. However, since SNNs rely on a large number of interconnects, electronics suffer from fundamental performance challenges due to resistor-capacitor parasitics and radiative effects in electronic links. As a result, power dissipation and inherent bandwidth-distance trade-offs limit the operational speed and bandwidth. In addition, although communication protocols such as address-event representation (AER) or time-division multiplexing (TDM) can attribute to an increase in computing capacity, they rely on transforming spikes into digital code instead of processing real-time physical pulses, resulting in loss of bandwidth and increasing latencies.[5]

Integrated photonics, on the other hand, is able to offer sub-nanosecond latencies, high bandwidth and low energy consumption due to the bosonic nature of photons. Photonics benefits from lower propagation loss in waveguides, lack of inductive and capacitive effects, and ability for parallelisation using wavelength division multiplexing (WDM) for increased capacity. Integrated two-section lasers comprising a gain and saturable absorber exhibit similar dynamics as observed in biological neurons, albeit on the nanosecond timescale. This raises the question of whether an ultra-fast and energy efficient optical SNN can be implemented in integrated photonics. The potential low-latency and high-speed capabilities could be beneficial in various applications, ranging from fundamental physics breakthroughs (e.g. high-energy-particle collision classification or nuclear fusion reactor plasma control), to non-linear programming (e.g. model predictive control (MPC) on hypersonic spacecraft), and optical signal processing in the GHz domain (e.g. wideband spectral sensing (WSS) for signal separation and localisation).[6–9]

Different strategies to achieve a photonic spiking laser neuron have already been proposed. Non-integrated versions such as an optically-pumped fibre laser comprising erbium and thulium doped fibres as gain and absorption sections or a graphene-based fibre laser showed excitability on a millisecond and microsecond timescales, respectively.[10,11] Examples of integrated approaches are cavity surface-emitting lasers (VCSELs)[12–14] or micro-disk lasers on a silicon-based platform[15]. However, due to the surface-emitting VCSELs and phase sensitive micro-disk laser, cascadability and integration with other components it remains challenging.[16] Another recent approach was based on a distributed-feedback (DFB) laser in active InP technology with a voltage-controlled saturable absorber.[17,18] This fully integrated, all-optical laser neuron showed self-spiking and excitable capabilities and cascability, at a relatively large threshold current of 90mA. Being an all-optical approach, it avoids any O/E/O conversion losses. However, as the cavity mirrors were formed by cleaving and applying coatings, only off-chip connections using fibres can be made, hindering also here future scalability to larger networks. The current state-of-the-art show that full integration of single neurons, cascading with subsequent neurons, or combining with other on-chip components for optical weighting remain challenging. A spiking neuron demonstration as part of a versatile and scalable photonic circuit platform would be needed to allow for future extension to complete networks.

We have focused on such a considerably more flexible approach and fabricated integrated laser neurons with stochastic self-spiking capabilities on an InP-based commercially-available generic technology platform.[19–21] This platform enables fast-prototyping and low-cost development of photonic integrated circuits (PICs) through the use of predefined building blocks and a foundry-specific process design kit (PDKs).[22–25] This enables PIC design at functional level, and targets a wide range of applications: e.g. tuneable wavelength lasers for gas sensing[26,27] and optical coherence tomography[28] (OCT), integrated optical phased arrays[29] for LiDAR, low-linewidth laser sources for quantum key distribution (QKD)[30], mmWave signal generation[31], and more. Also hybrid integration of this platform with a silicon-nitride platform (SiN) for increased complexity of PICs was demonstrated recently.[32] This vast portfolio of PICs demonstrates the strength and flexibility of generic integration. Recently, we have modelled and analysed the rich dynamics of a two-section integrated laser neuron under optical pulse injection in three dimensional phase space.[33–35] Depending on the gain current and saturable absorber voltage, the excitable, self-spiking, on-set or cw modes of operation were modelled. Moreover, the mirror reflectivity plays an important role in increasing the control parameter window for excitable operation.[34] Based on these findings, an optimised cavity design for an integrated two-section laser neuron was made.

Here, we report on the design and measurements of the first all-optical spiking laser neuron integrated on the building-block based InP generic technology platform. To the best of our knowledge, it marks the first demonstration of an integrated all-optical spiking neuron that is fully compatible with a mature, commercial PIC platform. This is an important step towards the realisation of a fully integrated all-optical SNN, since the platform allows to scale up and combine the laser neuron with any other photonic circuit. Combined with ring resonator weighting banks, it would yield a complete photonic SNN[13,36], and combined with a time-based learning algorithm such as spike-timing-dependent plasticity (STDP)[37,38], it could allow for online training.

In the next sections, the integrated two-section laser design, fabrication, and measurement results that demonstrate the excitable and self-pulsation modes will be discussed. Additionally, the injected and excited pulse energies, along with the wavelengths that trigger the device, will be determined, as these are important quantities for scaling towards fully integrated SNNs.

## 2. Results

The device was fabricated on a 4.0×4.6 mm$^2$ design cell and comprises a 220μm DBR, 70μm SA, 1mm SOA, and MIR, as shown in Fig. 1 (a), (c), and (d)-(e). This layout allows for external optical pulse injection though the DBR into the absorber, which will saturate and induce the excited response at the MIR output. The gain and saturable absorber lengths are a trade-off between the saturation energy and low cavity losses. The values are chosen such that high gain is expected to overcome the cavity losses, whereas absorption is strong enough to maximise the excitability operation window. A voltage-controlled phase shifter to locally change the refractive index was also fabricated, but not used in the experiments. Active regions are interconnected using shallow-etched waveguides, whereas the input waveguide connected to the DBR is made through a 180° bend deep-etched waveguide. Shallow-etched waveguides are connected to the laser's input and output mirrors and routed from the laser to the die edges. These waveguides could also be coupled to weighting components or other integrated optical laser neurons to form a cascaded network. The PIC was packaged and wedge-bonded to PCB breakout boards (Fig. 1 (b)), for easy access to the electrical contacts.

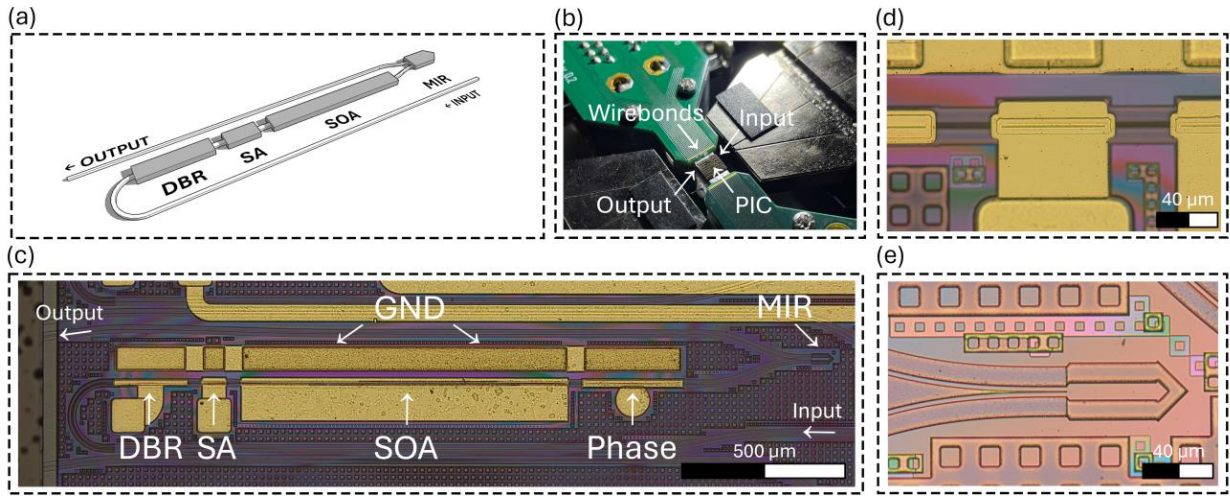

*Figure 1: Schematic overview of the laser (a), micrograph of the fabrication device on the generic integration platform before wedge-bonding (c), wedge-bonded die with electrical and optical connections for characterisation purposed (b), close-up micrograph of the saturable absorber and electrical isolation sections to the left and right, and the MIR in (d) and (e), respectively.*

2.1. Measurement setup to characterise the excitable laser neuron

To investigate the laser neuron's excitable operation mode, the measurement setup as shown in Fig. 1 was built. A femtosecond (fs) laser source (Menlo T-Light) was used to generate ultra-short (<90 fs) optical pulses at a repetition rate of 100MHz. The pulses were filtered by a 2 nm wide tuneable bandpass filter (Agiltron FOTF-040101333). This resulted in temporally broadened pulses with tuneable centre wavelength, which act as external spike input to the laser neuron. Using an autocorrelator, it was measured that the resulting FWHM was approximately 4 ps. An erbium doped fibre amplifier (Keopsys KPS-BT2-C-27-BO-FA) boosted the optical power, while a variable optical attenuator (JGR OA1) precisely controlled the injected power into the device. Next, the optical path was split into two arms. A reference arm (the dashed line in Fig. 2) connected to a real-time oscilloscope (LeCroy Wavemaster 8600A), while the second arm (the solid line in Fig. 2) directed the pulse train through a polarisation controller into the laser neuron. Since the absorption highly depends on the polarisation[39], the controller was set such that the generated saturable absorber photocurrent due to injected light was maximised. Using lensed single mode fibres, the external pulses were injected through the DBR into the saturable absorber, and light was collected at the MIR output. The output signal was amplified using an erbium-doped fibre amplifier (MPB 1RU-R35-CB) and filtered around the lasing wavelength using a 2 nm tuneable bandpass filter (JDS Fitel TB1500D) to block remnants of the injected pulses. It was then analysed using the oscilloscope and optical spectrum analyser (Yokogawa AQ6375). The average injected and output optical power were measured by two photodetectors (Agilent 81635A). The wedge-bonded PIC was mounted on an aluminium base and temperature-controlled to 15°C using a Peltier element and thermistor.

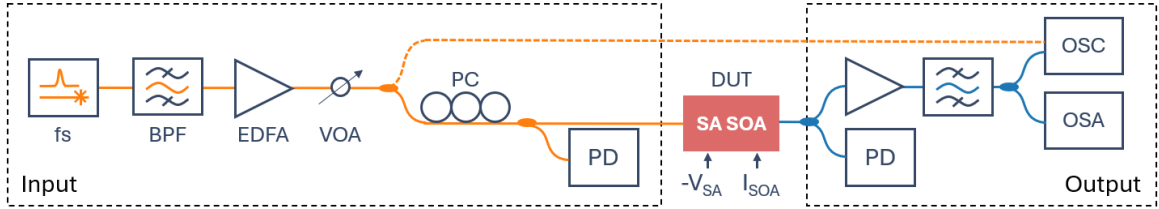

*Figure 2: Schematic overview of the measurement setup for excitability measurements. fs: femtosecond laser source, BPF: bandpass filter, EDFA: erbium-doped fibre amplifier, VOA: variable optical attenuator, PC: polarisation controller, PD: photodetector, DUT: device under test, OSC: real-time oscilloscope, OSA: optical spectrum analyser.*

2.2. Threshold current and spectrum

Using the setup as presented in the previous section, first the lasing threshold and spectrum are determined at an absorber-reverse bias voltage of 1.1V. In Fig. 2, the optical output power and the gain section voltage are plotted as a function of a increasing (thick solid) and decreasing (thin dashed) gain current sweep. Around 50mA a clear threshold and hysteresis are observed, which is expected in a two-section laser, and indicate the area of bi-stability.[40] At the threshold, the differential resistance is 11.9 Ω, whereas at higher currents (>60mA), the differential resistance is slightly lower at 10.7 Ω. In Fig. 4, the spectrum around the gain threshold current of 48.5mA is depicted. Above threshold, the laser operates in multimode and a number of cavity modes at a spacing of 154.2 pm are present, which is in agreement with the cavity length of approximately 2.08 mm at the lasing wavelength of 1547.12 nm.

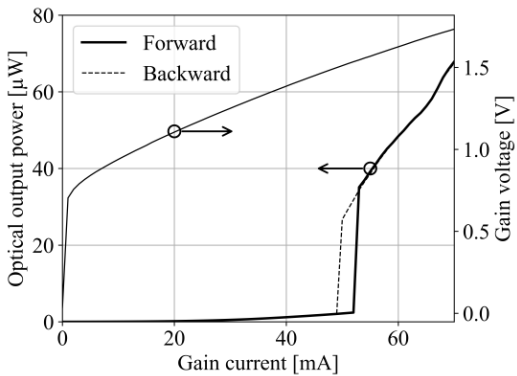
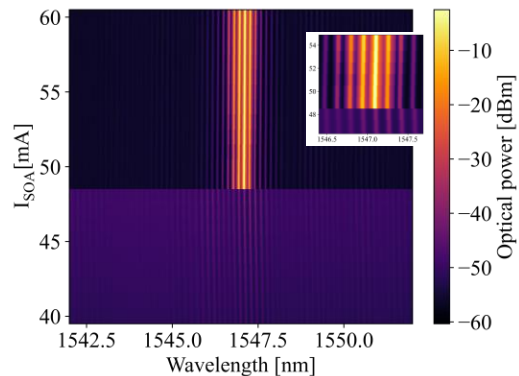

*Figure 3: Laser neuron's LIV measurement displaying the gain threshold current around 50mA and hysteresis at a reverse bias voltage of 1.1V.*

*Figure 4: Measured spectrum around the gain current threshold and reverse bias voltage of 1.1V. Inset: detailed spectrum showing the different cavity modes at a spacing of 154.2 pm.*

## 2.3. Excitability timetraces

From previous modelling work[33], it is known that excitable operation due to external pulse injection is expected slightly below the gain threshold current. If an external pulse of sufficient energy is injected into the saturable absorber, the laser produces an excited response in the form of an optical output pulse. By varying the injected pulse energy using the variable optical attenuator, this important feature was characterised. In the following experiment, optical pulses at a centre injection wavelength of 1547.0nm were injected into the laser, at a gain current of 52.0mA and a reverse bias voltage of 1.13V. As observed in Fig. 5(a), an injected pulse train with an time average optical power of -22.5dBm at the input lensed fibre results in a sustained train of excited output pulses with a FWHM of approximately 150ps (see inset). Gradually lowering the injected pulse energy results in the drop out of individual excited pulses, up until the point where no excited response was recorded, Fig. 5(b)-(d). The gradual decrease in number of excited responses is due to observed small pulse variations in the injected pulse train and possibly the presence of noise superimposed on the injected signal[41,42], resulting in individual pulses either laying above or below the excitability threshold.[11]

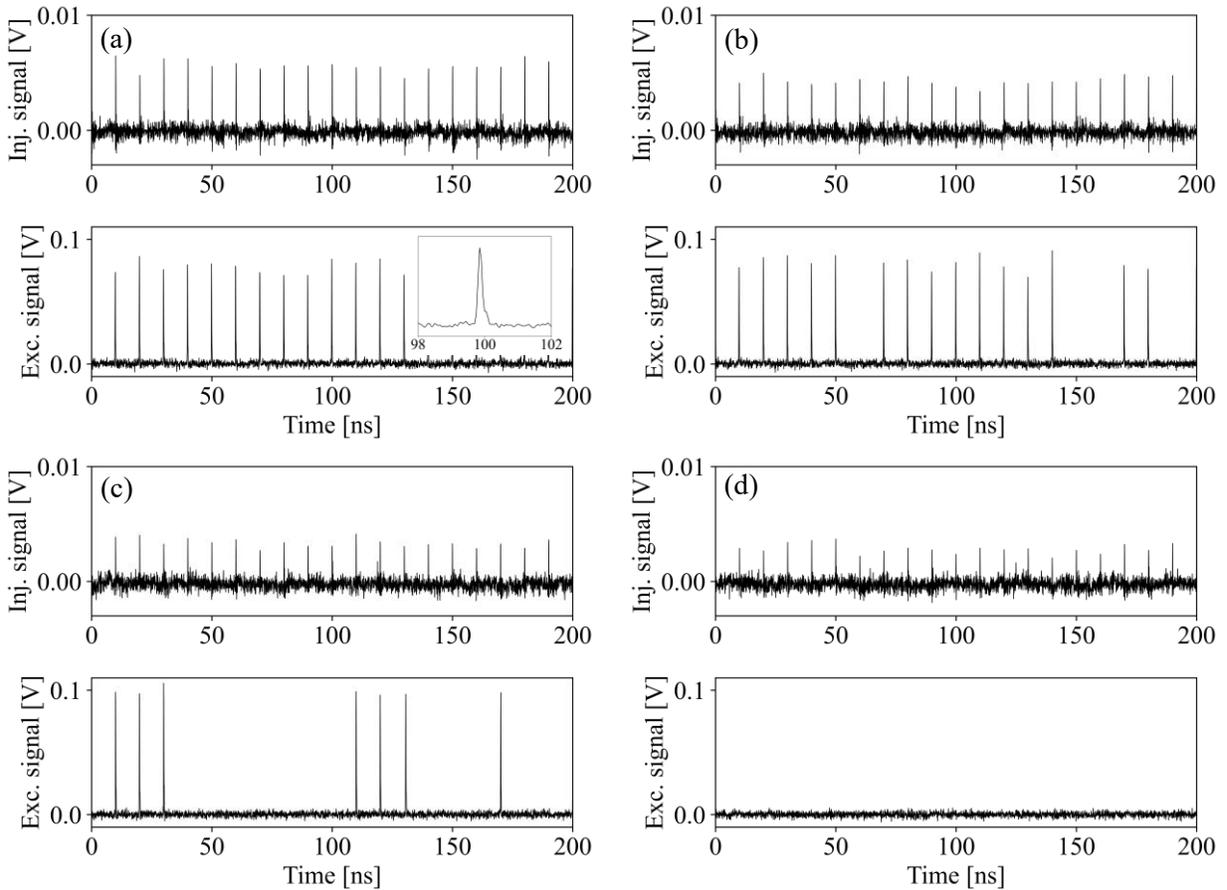

Figure 5: Four injected pulse trains at a fixed repetition rate of 100MHz, each with slightly lower injection strengths. The top figures in (a)-(d) show the injected signal, the bottom figures the response signal. Inset in (a) shows the excited pulse around 100ns.

## 2.4. Excited pulse energy and threshold

To quantify precisely the conditions for single-pulse excitation, the measured injected and output powers were used to calculate excited pulse energies. Fig. 6 shows the time average optical output power as a function of the calculated injected pulse energies for four different gain currents at a reverse bias voltage of 1.2V (left) and three different reverse bias voltages at a fixed gain current of 49.4mA (right). These conditions were chosen within the excitability region to study the relation of input and output pulse energies. By taking into account a realistic chip-to-fibre coupling loss of 3 dB for the InP generic technology platform, the injected pulse energies were calculated by dividing the measured average input power over the fixed repetition rate of 100MHz. In Fig. 6, for input pulse energies between approximately 10fJ/pulse to 50fJ/pulse, a threshold is observed. Below 10fJ/pulse, the injected pulse energy is not sufficient to induce an excitable response. For a relatively high input pulse energy of 65fJ/pulse, for all gain currents depicted in Fig. 6 (left), the injected pulses resulted in excitable responses. The output pulse energies range from 112.8 to 124.3fJ/pulse for gain currents of 49.0 and 50.2mA, respectively, when the input pulse energy is as little as 60fJ/pulse. The boundary where the input pulse energies are equal to the output pulse energies is marked by the black dashed line. To the left of this line, flattening of the average output power is observed. In this region, the injected pulse energy is lower than the output pulse energy. Thus, at a gain current of 50.2mA the minimum input pulse energy for 'full' excitability (i.e. all injected pulses induce response pulses) is approximately 33.2fJ/pulse, at a corresponding output pulse energy of 125fJ/pulse. Consequently, the laser neuron fully regenerates the input pulse and provides more than 5.7dB of gain.

When comparing the two figures, it becomes clear that the threshold steepness is not affected by the changes in the gain current or reverse bias voltage. However, an increase in the gain bias current not only shifts the excitatory threshold to a lower input pulse energy, it also increases the output pulse energy. Interestingly, lowering only the bias voltage shifts the excitatory threshold to higher pulse energies. Effectively, gain current and absorber voltage change the sensitivity of the laser neuron to the input pulse energies, as well as the output pulse power. These results undoubtedly show that the laser neuron is excitable. The important observation that the output pulse energy exceeds that of the input facilitates cascading of such laser neurons.

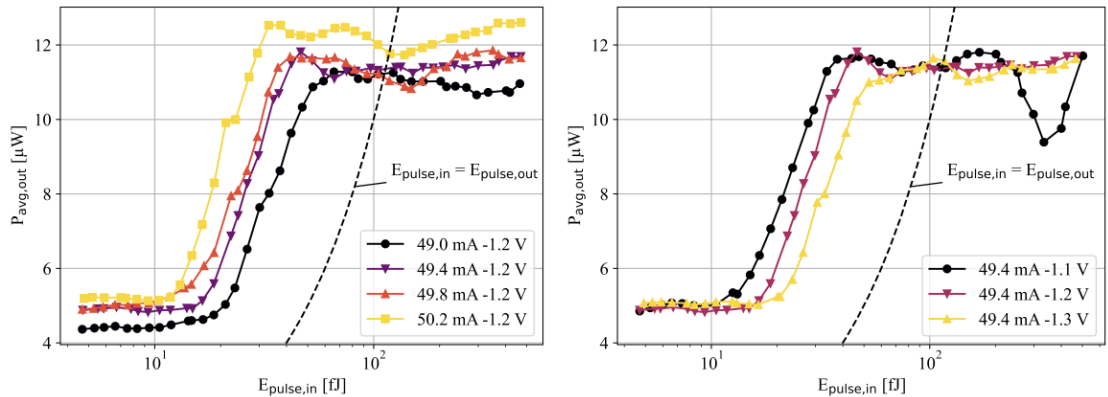

Figure 6: Average output power as a function of the injected pulse energy for four different currents (left), and three different voltages (right). The black dashed line shows the value at which the input pulse energy is equal to the output pulse energy, in case of excitable operation.

2.5. Excitatory delay

Another important property of a spiking laser neuron is the response latency or excitatory delay, i.e. the delay between an injected pulse and the excited response, which among others depends on the total injected pulse energy. Similar to the excitability measurements, the injected pulse energy was varied by the attenuator. Fig. 7 shows the injected pulses (top left) and the corresponding excited responses (bottom left). It is clear that lowering the energy of the injected pulse results in an increased time delay between the injected and the excited response. Note that due to the oscilloscope's limited bandwidth of 6GHz, oscillations in the tail of the injected pulses are observed. Since the injected pulse was measured to have a duration of approximately 4 ps, the oscillations are considered to be an artefact of the oscilloscope's limited bandwidth. By comparing the timings of the maxima of the injected and response peaks at 20 different input attenuation settings, the delay between the injected pulse and excited response was determined for three different gain currents (Fig. 7 (right)). It is observed that for relatively high energy pulse injection, the excitatory delay is small, i.e. below 100 ps. At smaller injected pulse energies, the excitatory delay increases to values above 220 ps. The effect of different gain currents on the delay is strongest when the input pulses are weakest (at 10 dB attenuation). In this case, the delay ranges from 320 to 234 ps for gain currents of 49.4 to 50.6mA, respectively. These are important findings, since they effectively show the sub-ns latency of the laser neuron and indicate ways to vary the delay period in a controlled manner for training algorithms such as STDP.[37]

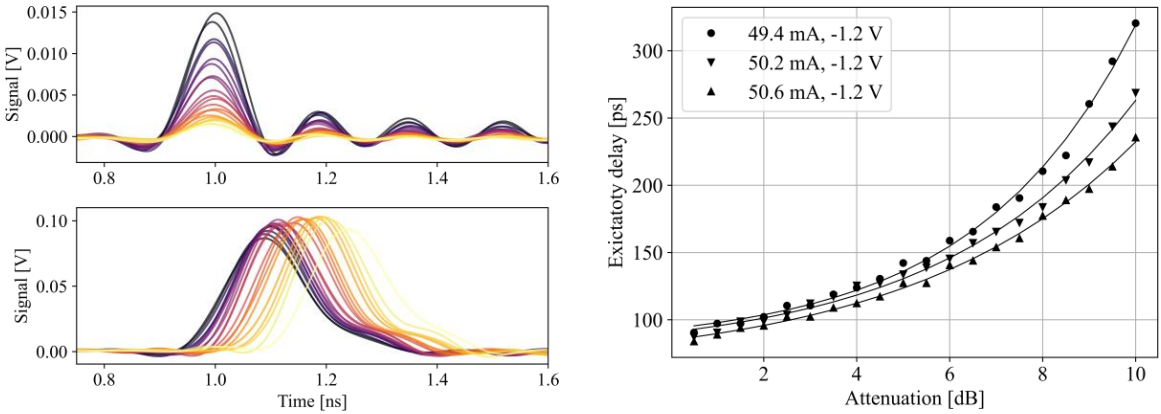

Figure 7: A total of 20 injected pulses with different amplitudes (top left), and their respective excitable responses (bottom left). The time delay between the input and output pulses for three different gain currents (right).

2.6. Dual pulse injection at the same wavelength

In biological neurons, the refractory period refers to the state after the action potential, in which the neuron cannot be excited by a second input perturbation (absolute refractory period), or a much stronger input perturbation is needed to elicit a second excited response (relative refractory period).[43] Effectively a dead time, it sets a boundary on the reaction time to multiple input pulses in short succession. To investigate this effect in the integrated laser neuron, the setup used for optical pulse injection was adapted such that two pulses with variable delay time are injected. Fig. 8 shows the schematic layout of the setup. After the variable optical attenuator, the signal is split into two arms. In one arm (the top arm in Fig. 8), a 300 ps tuneable optical delay line (OZ optics ODL-300) was placed to apply a variable pulse delay to that part of the pulse train. In the other arm, a second variable optical attenuator was placed to balance the optical powers in both arms. The two arms are merged and injected through the polarisation controller into the optical laser neuron.

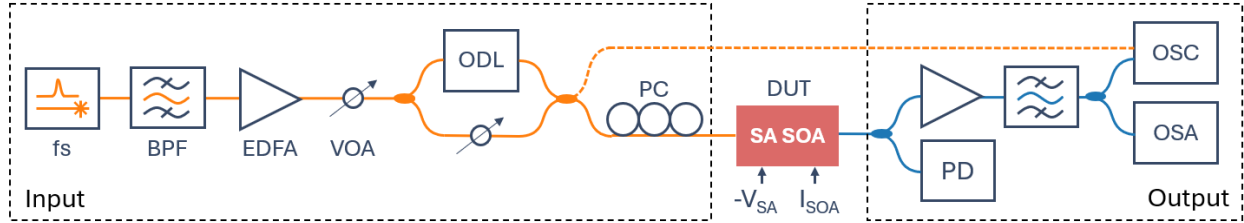

Figure 8: Schematic overview of the measurement setup for refractory period measurements. ODL: optical delay line.

Using the setup shown in Fig. 8, it is possible to inject two pulses with a variable pulse delay between the two pulses. Fig. 9 (left) shows the results for a gain current of 51.3mA and a reverse bias voltage of 1.20V, where 7 pulse pairs are injected, each with different delay between the two pulses, starting from 850ps for the first case to 1.15ns for the last case. At 1.06ns (the fourth case), a second excited response appears.

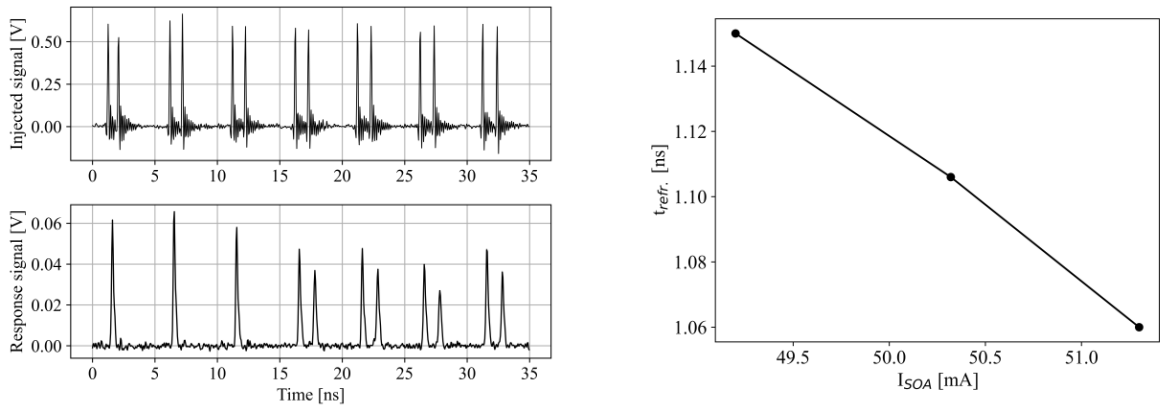

Figure 9: Seven injected dual pulse patterns with different time delays, starting from 850 ps up to 1.15ns (top left) and the excited responses (bottom left), time at which the second pulse in the excited response appears (right).

The refractory period depends on the biasing operation. In Fig. 9 (right), the refractory time versus three different gain currents is displayed, demonstrating shorter refractory times with increasing gain currents. The observed trend can be explained by considering the gain carrier concentration. It is well understood that after an excited response, the gain carrier concentration is recovering exponentially to reach its steady state value. The steady state value depends on the amount of injected current.[44,45] A higher injection current results in increased gain, while the steady-state losses remain unchanged. For a relatively high injection current, when a second pulse is injected during the recovery phase, the cavity losses can be overcome more easily, and thus a second excited response is formed earlier compared to the case when the injection current is low.

2.7. Dual pulse injection at different wavelengths

The widely discussed broadcast and weight protocol[36] exploits different wavelengths to scale up the number of synaptic connections, which requires the laser neuron to be susceptible to multiple input wavelengths. To demonstrate an excitable response to input signals at wavelengths other than 1547 nm, the previously explained setup (see Fig. 2) is extended. The femtosecond laser output is split into two different optical paths with unequal lengths, and an additional tuneable bandpass filter (filter 2, JDS Fitel TB1500D) is added in parallel to the already existing tuneable bandpass filter (filter 1, Agiltron FOTF-040101333), see Fig. 10. The different path lengths were achieved by adding extra fibre into one of the arms, and ensures there is no overlap in time between the two pulse trains. The setup now allows for dual pulse injection at different centre wavelengths, while it is still possible to resolve the two different input pulses in recorded timetraces.

In the following measurement, bandpass filter 1 and 2 are set to 1546.11 nm and 1544.77 nm, respectively, at . The gain current was set to 51.75mA, and the reverse bias voltage to 1.2V. These settings were chosen such that the spectral overlap between these signals was minimised, and the individual signals were able to excite the laser. Next, the dual pulse input signal was injected into the laser. Fig. 11 (left) shows the corresponding spectrum of the injected signal. It should be noted that the spectrum consists of a superposition of the amplified and bandpass-filtered signals, and that two different tuneable filters were used. This is the reason the spectrum shows different slopes at the edges of the spectrum. Nonetheless, it was verified that the optical power output in both arms was of equal strength with approximately equal response of the laser neuron at both wavelengths. In Fig. 11 (top right), the injected pulse train is shown, with the corresponding excited response in the bottom right figure. From the response signal, it is clear that both pulses at different wavelengths are able to excite the laser neuron. These results demonstrate that the laser neuron can be excited at two different injecting wavelengths which opens up the route to scale towards higher number of synaptic connections through wavelength division multiplexing (WDM).

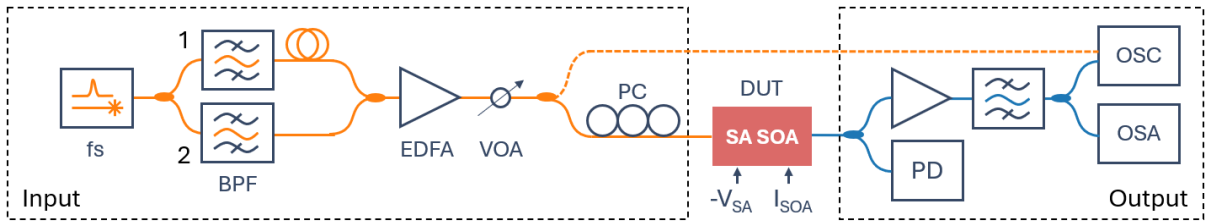

Figure 10: Schematic overview of the measurement setup for dual pulse injections measurements.

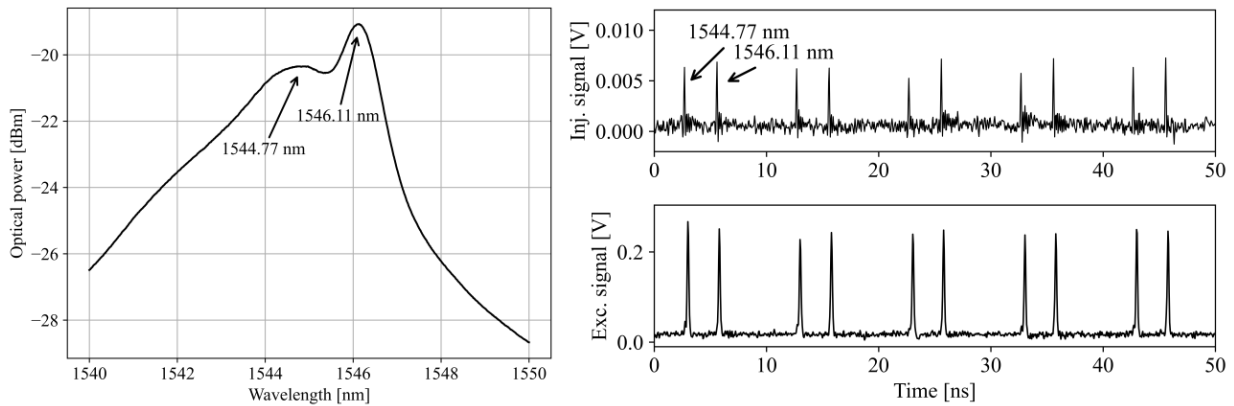

Figure 11: Spectrum of the injected light showing the two distinct injection wavelengths (left). Timetrace of the injected light (top right) and the laser neuron response to these inputs (bottom right).

2.8. Injection wavelengths

The results on dual pulse injection at different wavelengths demonstrate that excitability is not limited to pulse injection at only one single wavelength. To explore the potential of injecting pulses at different injection wavelengths, the bandpass filter of the setup shown in Fig. 1 was used to sweep the injection wavelength centre wavelength of a single pulse from 1543.5 to 1547.0nm. This is the wavelength region directly below the lasing wavelength, as presented before in Fig. 4. The upper bound of this range was chosen due to the 2 nm wide stopband of the integrated DBR mirror, which suppressed external pulse injection at wavelength above 1547.0nm. Moreover, due to saturable absorber material properties, at wavelengths above the lasing wavelength the absorption effect reduces significantly, and thus the excitable character vanishes.[39] The colormaps in Fig. 12 (a)-(b) show the excitable response ratio, i.e. the ratio between the number of excited pulses and the number of injected pulses, as a function of the injected centre wavelength and gain current for reverse bias voltages of 1.2 and 1.3V, respectively. In these colormaps, a value of '1' (yellow) means all injected pulses results in an excited response, whereas a value of '0' (black) means no injected pulse resulted in an excited response. From these figures it is observed that for a relatively large range of wavelengths, an excitable response value of '1' was recorded. This means that over this entire wavelength range, injected pulses always result in an excited response. For the combination of shorter wavelengths and lower currents, there exists a region where injected pulses that result in an excited response is not stable anymore, i.e. excitation fails. The boundary at which this takes place shifts by changing the reverse bias voltage, and is indicated by the dashed lines in Fig. 12 (a)-(b). The largest range of wavelength over which excitability is successful (i.e. the ratio is 1), is at a relatively low reverse bias voltage and high gain current. From the measurements, a window of 3 nm can be obtained.

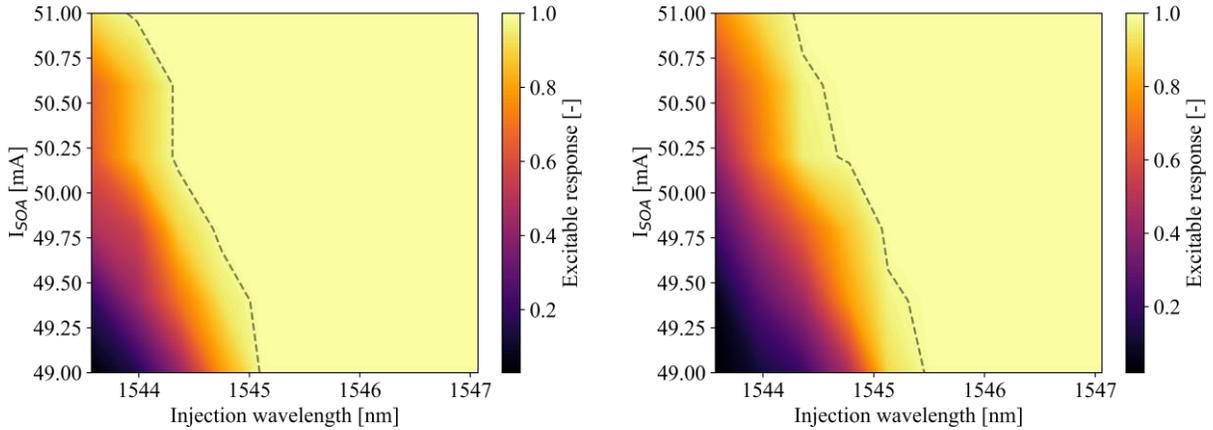

*Figure 12: Excitable response measurements for reverse bias voltages of 1.2 (left) and 1.3V (right). The dashed line shows the boundary of fully successful excitation, i.e. all injected pulses result in excitable responses vs. a gradual slope where not all injected pulses result in an excitable response.*

2.9. Measurements of a self-pulsating laser neuron

In addition to the excitable operation mode, the laser neuron exhibits a different mode of operation. Previously simulated laser neurons on the generic platform show a Q-switching mode, or self-pulsating mode, for sufficiently strong absorber reverse bias voltages and high gain currents.[34] In this mode, a sustained train of pulses with a distinct pulsating frequency is generated without any external stimulus, which in the context of biological neurons is analogue to, for example, a train of pulses with a fixed inter spike interval (ISI) of muscle motor neurons[46] or ganglion cells[47] in the eye. To study the properties of this self-pulsating mode, such as the frequency as well as the pulse energy, the setup as used in the excitability measurements was used (Fig. 2). In this case, however, the gain current was swept from 40mA (below threshold) to 100mA, in steps of 1mA, while for every increment of the gain current, the absorber reverse bias voltage was swept from 0 to 3.5V in steps of 50 mV. In each voltage step, the oscilloscope was used to capture a timetrace over a time duration of 1000ns, while the photodetector measured the time average optical output power. From the recorded timetraces, the frequency of the pulsating output was determined. This wide range of gain currents and absorber bias voltages covers not only the excitable operation mode, but also allows for capturing pulses at any other repetition rate that are generated due to Q-switching. Fig. 13 (left) shows the 2D parameter space and the analysed output frequency mapped onto the colour scale. The area where the analysed output frequency was 100MHz, which was due to the injected pulses, is marked as the excitable operation regime. Next to this area, the self-pulsating regime is marked. In this region, output frequencies ranging from 100MHz up until 1.1GHz were recorded. Note that in these results, only the frequency of a pulsating signal is considered. From previous simulations on a laser neuron in this technology platform, it is expected that for sufficiently large gain currents and low reverse bias voltages (top right in Fig. 13 (left)), the laser operates in the cw regime.[34,48]

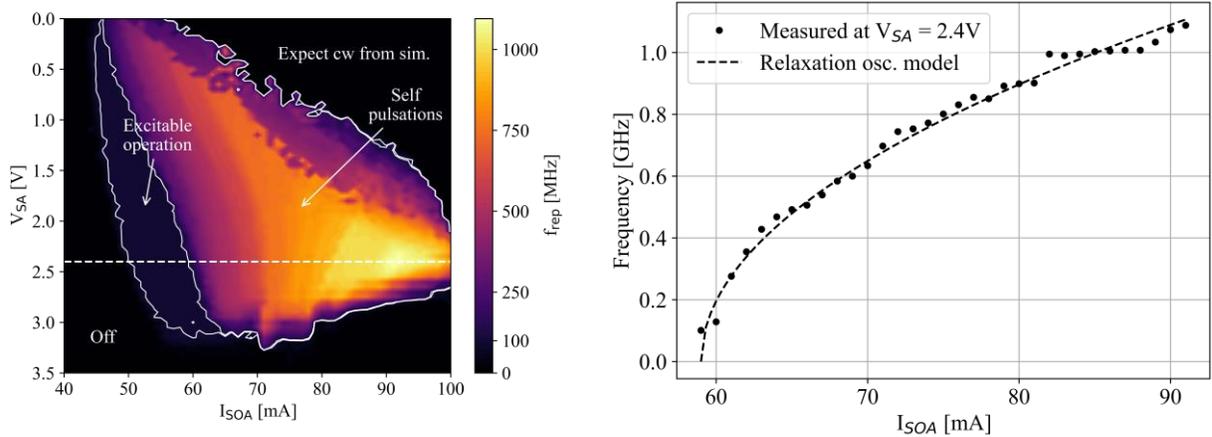

Figure 13: Analysed frequency of the pulsed laser output for gain current between 40 and 100 mA and reverse bias voltages between 0 and 3.5V (left). Frequencies along the dashed line, at a reverse bias voltage of 2.4V, are shown in more detail on the right, where the frequencies are fitted to the relaxation oscillation model for a semiconductor laser.

The dashed line in Fig. 13 (left) marks a reverse bias voltage of 2.4V, of which the recorded frequencies above the lasing threshold of 59.0mA are shown in Fig. 13 (right). At this specific reverse bias voltage, the maximum frequency of 1.1GHz was measured. Based on the values presented in Table 1, which are matched to previously measured data on the InP generic platform, the measured frequencies were fitted to the following expression for the relaxation oscillation frequency $\omega_r$ for a semiconductor laser operating above threshold:[49]

$$\omega_\text{r} = \sqrt{\frac{\Gamma v_\text{g} a}{qV} \eta_\text{i} (I - I_\text{th})} \tag{1}$$

Table 1: Values used for fitting the measured frequencies in Fig. 13

| | | | | | |
|---|---|---|---|---|---|
| $\Gamma$ | optical confinement factor [50] | 0.053 [-] | $V$ | volume of the active region | $2.14 \cdot 10^{-16}$ m$^3$ |
| $v_\text{g}$ | group velocity [51] | $8.14 \cdot 10^7$ ms$^{-1}$ | $\eta_\text{i}$ | injection efficiency | 0.6 [-] |
| $a$ | differential gain | $2 \cdot 10^{-20}$ [-] | $I$ | gain current | variable [mA] |
| $q$ | elementary charge | $1.60 \cdot 10^{-19}$ C | $I_\text{th}$ | threshold current | 59mA |

From Fig. 13 (right) it is clear that the model and the measured data are in good agreement, and that the measured frequencies can be very well approximated by a square root characteristic.

From the measured time-averaged optical output power and the measured pulsation frequency, an estimation for the energy per pulse was derived for the laser neuron output spikes. At an absorber reverse bias voltage of 1.1, 1.2, and 1.3V the pulse energies are 123, 104, and 93fJ/pulse, respectively, which in all three cases is above the excitability threshold depicted in Fig. 6. This is an important observation, since it is strong evidence that the generated pulses are of sufficient energy to excite a subsequent laser neuron operating in the excitable regime. In addition, it should be noted that the pulsation frequency depends on both the reverse bias voltage, as well as the gain current. By changing either of these quantities, the self-pulsation frequency can be tuned to a desired value, and thus it can be considered to be an electronic-to-photonic spike rate encoder.[52]

## 3. Discussion

The results presented in this paper demonstrate the first all-optical integrated laser neuron realised on an InP generic integration technology platform and shows important features, such as excitability, refractory period, dual pulse and multi-wavelength injection and self-spiking. The refractory period for the proposed device is measured to be approximately 1.10ns. During the experiments, it can be noted that at the boundary of the refractory period, the two excited pulses have different amplitudes when compared to the case of single pulse excitation (Fig. 9). This effect was measured before in VCSEL-based optical neurons and is attributed to the interaction of the rising front of the second injected pulse with the generated excited pulse.[53]

Another aspect to consider is the total energy consumption of a single laser neuron. The lowest lasing threshold for the proposed device is approximately 50mA at an absorber reverse bias voltage of 1.1V (Fig. 2). Although, substantially larger than a packaged and commercially-available VCSEL-based approach,[12] it is almost a factor of two lower compared to the integrated DFB-based laser neuron (90mA)[17], a significant improvement. Based on the gain current, the differential resistance (Fig. 2) and the self-pulsation frequency (Fig. 13), the consumed electrical energy in the self-pulsating mode can be translated into energy-per spike operation (SOP), which varies from 0.13 nJ/SOP at 230MHz down to 67.0pJ/SOP at 730MHz. An estimate of the lowest energy consumption in the excitable regime can be made based on the refractory period of approximately 1.0ns (Fig. 9), and the lowest possible gain current and absorber biasing voltage at which excitable operation is still possible (46mA and 0V (Fig. 13)), which results in an energy consumption of 25.2pJ/SOP. It must be noted that the energy per pulse can be further lowered by optimising the overall cavity losses. Aside from process-related reduction of waveguide losses, this can be done by increasing the DBR reflectivity, or reducing the overall cavity length. The latter is possible by reducing the gain section length, or by reducing the passive waveguide lengths that connects the individual building blocks. Another method to reduce the energy per spike operation is to further decrease the refractory period by reducing the gain and saturable absorber carrier lifetimes through optimisation of the active layer stack. It is worth mentioning that the laser design itself is essentially a trade-off between performance and footprint, since, for example, increasing the DBR reflectivity or decreasing the gain section length eventually also decreases the optical output power, which has an impact on overall cascadability.

The multi-wavelength injection results show a bandwidth of 3 nm in which the neuron can operate in the excitable state. This allows for multiple synaptic connections to other neurons following the broadcast and weight protocol. If pulses are generated from preceding laser neurons with a pulse duration of 100 ps and a transform limited bandwidth of 25 pm, the available optical window can support up to 120 wavelengths, a sufficiently large number of connections for small-scale problems.

Yet another point to consider is the size of the integrated laser, which for the proposed device is approximately $0.38 \times 2.4$ mm$^2$ (height × width). The InP generic integration technology platform offers a standard design area of $4.0 \times 4.6$ mm$^2$ for a single cell, or alternatively, $8.0 \times 4.6$ mm$^2$ for a double cell. This leads to a maximum number of 19 or 39 neurons in a staggered layout on a single or double cell, respectively. These estimations are based on the current design (Fig. 1), however, by replacing the relatively large gold DC probing pads by smaller electronic wirebonding pads, and by replacing the 180° bend waveguide by a straight connection, a substantial reduction in the height to <80μm is possible. In addition, an improved cavity with a shorter gain section and shorter passive connections between building blocks, could lead to a 50% reduction and a total footprint reduction of around 10 times, without changing the PIC platform.

As mentioned in the introduction, the InP generic integration platform allows for designing photonic integrated circuit with a high degree of complexity by combining active and passive building blocks through integrated waveguides. This opens up the opportunity to integrate the proposed laser neuron with tuneable ring resonators[13,36], or, alternatively, with Mach-Zehnder interferometers through the use of building blocks such as N×N multimode interference (MMI) couplers, wavelength-selective arrayed-waveguide gratings (AWG), and waveguide crossings. In addition to passive components, reverse bias-controlled electro-optic effects can be exploited to implement efficient and fast tuneable ring resonators.[54] Combining such components on-chip with the proposed device will yield reconfigurable weighting capabilities. Previous work on VCSEL-based neurons demonstrated that a small hybrid photonic-software network comprising 4 parallel laser neurons detected and tracked a single feature in a sequence of input images. By combining two separate layers of 3 and 5 neurons with an offline STDP learning algorithm, classification of hand-written inputs from the MNIST dataset was demonstrated.[55] Hence, using a small number of neurons, classification of data is already possible. Here, we have shown that the spiking neuron function can be realised all-optically and fully integrated in a generic InP platform, paving the way towards densely-connected, multi-wavelength optical SNNs on a single photonic chip, a significant step forward.

To conclude, this work provides the first experimental demonstration of an all-optical excitable laser neuron integrated on the active-passive InP generic technology platform, which is a generic foundry service with an extensive library of mature building blocks. The proposed device comprises a 220μm distributed Bragg reflector, a 70μm saturable absorber, a 1 mm semiconductor amplifier, and an multimode interference reflector. At a lasing threshold of approximately 50mA, the device is capable of generating an ultra-fast excitable response pulse (i.e. width ~100 ps) of 112 to 124fJ/pulse when an input pulse of only a 60fJ/pulse is injected into the saturable absorber. Depending on the saturable absorber and gain current biasing operation points, the sensitivity to the input energy can be tuned, or the output pulse energy can be tuned, respectively. The refractory time was measured to gain current dependent and around 1.1ns. Furthermore, optical pulse injection up to 3 nm below the lasing wavelength was demonstrated, which opens up the way to wavelength multiplexing based synaptic connections. In addition to the excitable operation mode, a self-pulsating mode with a tuneable output frequency between 100MHz up until 1.1GHz for pulse energies on the order of 100fJ/pulse are measured. These extensive measurement results indicate the proposed laser can be used as an integrated all-optical laser neuron and may be further integrated into a photonic spiking neural network.

## 4. Materials and Methods

The photonic laser neuron is implemented on an InP generic platform that offers various building blocks which can be combined to more complex components or circuits.[19,20] For optical gain and saturable absorption, quantum wells-based semiconductor optical amplifiers (SOA) and saturable absorbers (SA) are grown using metal organic chemical vapor deposition (MOCVD). The active regions are integrated with bulk InGaAsP passive waveguides for low-loss transmission through butt-joined regrowth technology. This allows for flexible positioning of active and passive regions with low coupling loss (~0.1dB) and low reflections (<40dB).[19] Distributed Bragg reflectors (DBRs) are patterned on waveguides using ArF lithography, of which the reflectivity and wavelength can be tuned in design.[23,56] Alternatively, a broadband multimode interference reflector (MIR) provides a fixed reflectivity.[22,57] Components can be connected using 2.0μm wide shallow-etched rib waveguides for low-loss connections, or 1.5μm deep ridge waveguides for sharper bend radii. To electrically disconnect active regions, electrical isolation sections are formed between the gain and absorber by selectively removing the conductive p-doped cladding layer. To improve uniformity during planarisation, additional dummy loading structures are added to the empty design areas. After full processing, individual cells are diced and undergo anti-reflection coating.

**Data availability statements**

The data that support the findings of this study are openly available at the following URL/DOI: 10.4121/d0858cf5-a236-40fa-82d6-2ecfd8b7e88c.

**Acknowledgements**

The authors would like to acknowledge Nazca Design, which was used to generate the mask layout in this work, SMART Photonics, Eindhoven, for fabricating the device that was measured and presented in this paper, as well as the optical laboratory (OLA) at the Eindhoven University of Technology for providing the resources to perform the experiments.

## Author Contributions

Lukas Puts: designed PIC layout, interpreted results and drafted manuscript, Daan Lenstra: supported interpretation and analysis of results, revised manuscript, Kevin A. Williams: providing resources at the Eindhoven University of Technology, Weiming Yao: supported PIC design layout, interpretation and analysis of results, revised manuscript.

## Conflict of Interest Statement

The authors declare no conflict of interest.

## ORCID iDs

Lukas Puts https://orcid.org/0000-0001-5416-3006
Daan Lenstra https://orcid.org/0000-0002-4000-8897
Kevin Williams https://orcid.org/0000-0001-9698-9260
Weiming Yao https://orcid.org/0000-0002-4558-317X